\documentstyle[prd,aps,epsfig,
tighten,
multicol]{revtex}
\begin{document}
\preprint{KIAS-P98001 \hspace{.3cm}  hep-ph/9802209}

\title{Three Neutrino $\bf \Delta m^2$ Scales and Singular Seesaw Mechanism}

\author{E. J. Chun$^{a}$, C. W. Kim$^{a,b}$, and U. W. Lee$^{c}$}

\address{ $^{a}$ Korea Institute for Advanced Study,
                207-43 Cheongryangri-dong, Dongdaemun-gu,
                Seoul 130-012, Korea}
\address{ $^{b}$ Department of Physics \& Astronomy, 
                The Johns Hopkins University,
                Baltimore, MD 21218, USA }
\address{ $^{c}$ Department of Physics,
                Mokpo National University,
                Chonnam 534-729, Korea }
\address{Email addresses: ejchun@kiasph.kaist.ac.kr, 
cwkim@kias.kaist.ac.kr, uwlee@chungkye.mokpo.ac.kr}  
%
\maketitle
\begin{abstract}
It is shown that the singular seesaw mechanism can simultaneously
explain all the existing data supporting nonzero neutrino 
masses and mixing.  The three mass-squared differences that are 
needed to accommodate the atmospheric neutrino data 
(through $\nu_\mu - \nu_s$ oscillation), 
the solar neutrino data via MSW mechanism  
(through $\nu_e - \nu_\tau$ oscillation), 
and the positive result of $\nu_\mu - \nu_e$ oscillation from LSND
can be generated by this mechanism, whereas the vacuum oscillation 
solution to the solar neutrino problem is disfavored.
We find that the electron and tau neutrino masses are of order
$10^{-3}$ eV, and the muon neutrino and a sterile neutrino 
are almost maximally mixed to give a mass of order 1 eV.
Two heavy sterile neutrinos have a mass of order 1 keV
which can be obtained by the double seesaw mechanism with an intermediate
mass scale $\sim 10^5$ GeV.  A possible origin of such a scale is discussed.
\end{abstract}
\begin{multicols}{2}
%
\section{Introduction }
%

There are several neutrino oscillation experiments 
which have indicated  the nonzero neutrino masses and mixing.
The solar neutrino problem is the first to be noted.
The deficit of the solar neutrinos
predicted by the Standard Solar Model (SSM) \cite{SSM}
can be explained by neutrino oscillation between
$\nu_e$ and $\nu_x$.
The $\nu_x$ can be $ \nu_\mu \,, \ \nu_\tau $, 
or a sterile neutrino.
In the case of resonant MSW transitions \cite{MSW},
it was found \cite{SOLMSW}
that the oscillation parameters
$\Delta{m}^2$ and $\sin^{2}2\vartheta$ ($\vartheta$ is the mixing angle)
given by
\begin{eqnarray}
3 \times 10^{-6} \lesssim &\Delta{m}^{2}_{\rm solar}& 
 \lesssim 1.2 \times 10^{-5} \nonumber \\
4 \times 10^{-3} \lesssim &\sin^{2}2\vartheta_{ex}& 
  \lesssim 1.2 \times 10^{-2} 
\end{eqnarray}
can explain the solar neutrino problem.
The solar neutrino problem can also be solved by invoking vacuum
neutrino oscillations \cite{barger},
in which case the neutrino mass-squared difference is
about $\Delta m^2 \sim 10^{-10} \mathrm{eV}^2$.

Another hint of the neutrino masses and mixing comes from
the experiments on the atmospheric neutrinos.
The indications in favor of
$ \stackrel{\makebox[0pt][l]
{$\hskip-3pt\scriptscriptstyle(-)$}}
{\nu_{\mu}}
\to
\stackrel{\makebox[0pt][l]
{$\hskip-3pt\scriptscriptstyle(-)$}}
{\nu_{x}} $
oscillations ($x\neq\mu$) have been found in the
Kamiokande \cite{Kamiokande-atmospheric}, IMB \cite{IMB}, and recent 
Super-Kamiokande \cite{SKam}, Soudan-II \cite{SoudanII}
atmospheric neutrino experiments.
{}From the analysis of the Super-Kamiokande and the other data
the allowed ranges for the oscillation parameters were obtained
\cite{SKam,newatm}, assuming $\nu_{\mu} \leftrightarrow \nu_{\tau}$,
\begin{eqnarray}
3 \times 10^{-4} \lesssim &\Delta m^2_{\rm atm}& \lesssim 7 \times 10^{-3} 
\nonumber \\
0.8 \lesssim &\sin^2 2\vartheta_{\mu\tau}& \lesssim 1 \ .
\end{eqnarray}
The recent results from the Chooz reactor experiment appear to 
exclude the $\nu_\mu \leftrightarrow \nu_e$ oscillation as a solution to the 
atmospheric neutrino problem \cite{Chooz,SKam}.
The neutrino oscillations 
$ \nu_{\mu} \leftrightarrow \nu_{\tau} $ and  
$ \nu_{\mu} \leftrightarrow \nu_{s} $
give rise to a similar result for the atmospheric neutrino anomaly
since $\nu_s$ and $\nu_\tau$ are not distinguishable in the
current experiments.
New experiments with high statistics in Super-Kamiokande and ICARUS
can provide the discrimination by measuring the neutral current 
events appearing, for instance, in neutral pion productions \cite{VS}.

Finally, indications
in favor of $ \bar\nu_\mu \to \bar\nu_e $ oscillations
have been found recently in the LSND experiment \cite{LSND},
in which antineutrinos originating from the decays 
of $\mu^+$'s at rest were detected. The KARMEN experiment \cite{KARMEN}
will be able to crosscheck the positive result of LSND in the near future.
{}From the analysis of the data of the LSND experiment
and the negative results of other short-baseline experiments
(in particular, the Bugey \cite{Bugey95} and 
BNL E776 \cite{BNLE776} experiments),
one obtains the oscillation parameters: 
\begin{eqnarray}
 0.3 \lesssim &\Delta{m}^2_{\rm LSND}& \lesssim 2.2 \, \mathrm{eV}^2 
\nonumber \\
10^{-3} \lesssim &\sin^2 2\vartheta_{e\mu}& \lesssim 4 \times 10^{-2} \,.
\end{eqnarray}

All the existing neutrino experiments which are in favor of 
neutrino oscillations can not be accommodated by a scheme 
with mixing of ordinary three neutrinos.
That is, the allowed range of the mass-squared differences 
which explain the solar neutrino, atmospheric neutrino, 
and LSND neutrino experiments do not overlap at all.
In order to obtain three independent mass-squared differences 
which can explain the known three experiments, 
we need at least four massive neutrinos \cite{four,five,BGKP,BGG96,OY96}.
Furthermore, 
only two schemes of neutrino mass-squared differences 
are compatible with the results of all the experiments \cite{BGG96}.
Four neutrino masses are divided into two pairs
of almost degenerate masses separated by a gap of
$ \sim 1 \, \mathrm{eV} $
which is indicated by the result of the LSND experiments:
\begin{eqnarray}
&& \mbox{(A)}
\qquad \underbrace{ \overbrace{m_1 < m_2}^{\mathrm{atm}}
                \quad \ll \quad
                    \overbrace{m_3 < m_4}^{\mathrm{solar}}
                  }_{\mathrm{LSND}} \,, 
\nonumber     \\[-2mm]
&& \label{AB} \\[-2mm]
&& \mbox{(B)}
\qquad \underbrace{ \overbrace{m_1 < m_2}^{\mathrm{solar}}
                \quad \ll \quad
                    \overbrace{m_3 < m_4}^{\mathrm{atm}}
                  }_{\mathrm{LSND}} \;. 
\nonumber
\end{eqnarray}
In (A),
$\Delta{m}^{2}_{21}$ 
is relevant for the explanation 
of the atmospheric neutrino anomaly
and
$\Delta{m}^{2}_{43}$ 
is relevant for the suppression 
of solar $\nu_e$'s.
In (B),
the roles of $\Delta{m}^{2}_{21}$
and $\Delta{m}^{2}_{43}$ are interchanged.

If there exists a fourth neutrino, it has to be  a sterile neutrino as 
indicated by the invisible decay width of $Z^0$.
The mixing between a fourth neutrino and active neutrinos could be
constrained by the big-bang nucleosynthesis.
The active--sterile neutrino mixings suggested by the known
neutrino experiments may increase the effective 
number of neutrinos and deplete the electron neutrino and antineutrino
populations in the nucleosynthesis era, and thus alter significantly
the prediction for the primordial $^4$He mass fraction.
Formerly, the effective number of neutrino species was considered to be
less than four, and therefore  
the large mixings between a sterile neutrino and the active neutrinos
solving the solar neutrino and the atmospheric neutrino problem 
were disfavored \cite{NOATMS}.
However, due to the recent observational determinations 
of the primordial deuterium abundance\cite{DH,DHa},
the big-bang nucleosynthesis constraints on the 
non-standard neutrinos have been revised. 
It has been argued that the effective number of neutrino species more 
than four (but less than five) can be acceptable \cite{NNUG4},
in which case there is room to bring one  sterile neutrino
species into equilibrium with the known neutrinos in 
the early universe and therefore no constraints on 
active--sterile neutrino oscillation parameters can be drawn.
More interestingly, it is possible to reconcile sterile neutrinos with
big-bang nucleosynthesis  even if the effective number of neutrino species
turns out to be less than three.  It has been found that
the active--sterile neutrino oscillation solving the 
solar or atmospheric neutrino problem
does not increase the effective number of neutrinos  
if the relic neutrino asymmetry
is large enough ($L_\nu > 10^{-5}$) \cite{FV}. 

In the conventional seesaw mechanism, right-handed neutrinos are heavy 
to make active neutrinos very light.  In this paper, we explore the 
possibility that a right-handed neutrino remain light, that is, the 
right handed-neutrino mass matrix is singular and has rank two in a three
generation model.  This is called the singular seesaw mechanism \cite{GLASHOW}
which will be analyzed in section II.  Three mass-squared differences for the
light four neutrinos will be parameterized in terms of two mass scales; 
the Dirac mass and Majorana mass of heavy right-handed neutrinos.
By construction, the sterile neutrino (a light right-handed neutrino)
and an active neutrino are maximally mixed, which is desirable for 
the solution of the atmospheric neutrino anomaly. 
In section III, we will find the region of the two mass parameters by which
all the neutrino data mentioned above can be accommodated.
It is then required that the Dirac mass scale is $\sim 1$ eV and the heavy
Majorana mass scale is $\sim 1$ keV.  Such low mass scales are shown to 
imply the presence of an intermediate scale $\sim 10^5$ GeV and the grand
unification scale $\sim 10^{16}$ GeV in the double seesaw mechanism 
introduced in section IV. Finally, we conclude in section  V.

\section{Singular seesaw mechanism}

In the singular seesaw mechanism \cite{GLASHOW},
the neutrino mass matrix is written by
\begin{equation} \label{firstM}
{\cal M} = \left[ \matrix{ 0 & \epsilon M_D \cr 
                 \epsilon M_D^{\dag} & M_M }
   \right]
\end{equation}
where $M_D$ is the usual Dirac neutrino mass matrix and $M_M$ is
the right-handed Majorana neutrino mass matrix 
taken to be a singular (rank-two) $3 \times 3$ matrix.
Here, we assume that there is no hierarchical structure in the mass matrices
$M_D$ and $M_M$ whose elements are of the same order of magnitude, denoted by 
$M$.  The  small number $\epsilon$ encodes the hierarchical structure 
of the Dirac and sterile (right-handed) neutrino masses.  
The value $M$ is related to physics of lepton number 
violation and new physics.
Two parameters $\epsilon$ and $M$ are to be determined later.
We write the neutrino states in the interaction basis as 
\begin{equation}
\Psi_I = \left[ \matrix{ \psi_{(e , \mu , \tau), l} \cr
                                \psi_{(e , \mu , \tau), S} }
         \right]
\end{equation}
where 
$ \psi_{(e , \mu , \tau), l} $ and  $ \psi_{(e , \mu , \tau), S} $ 
represent, respectively, the three standard active and sterile neutrinos. 
In the context of the singular seesaw mechanism, 
a combination of 
$ \psi_{(e , \mu , \tau), S} $ becomes light.
In order to obtain physical neutrino states and mass eigenvalues,
we perform diagonalization by several steps.
First step is to diagonalize the Majorana part 
by a rotation matrix $R$ such as 
\begin{equation}
{\cal M} =
\left[  \matrix{ 1 & 0 \cr 0 & R^{T} }
\right]
\left[
        \matrix{          
        0 & \epsilon M_D R^{T} \cr 
        \epsilon RM_D^{\dag} & \tilde{M}_M
        }
  \right]
\left[
        \matrix{ 1 & 0 \cr 0 & R }
\right]
\end{equation}
where
\begin{eqnarray}
&& \tilde{M}_M = R M_M R^{T} = {\mathrm \  diagonal \  matrix} \,.
\label{tilmm}
\end{eqnarray}
Since $M_M$ is a rank-two singular matrix, 
we can take zero for the $(11)$-element 
of the diagonal matrix $ \tilde{M}_M$.
Then, 
we rewrite the mass matrix as
\begin{equation}
\left[
        \matrix{
        0 & \epsilon M_D R^{T} \cr
        \epsilon RM_D^{\dag} & \tilde{M}_M
        }
  \right]
=
\left[
        \matrix{
        \epsilon M_\alpha  & \epsilon M_\beta \cr 
        \epsilon M_\beta^{\dag} & \tilde{M}_m
        }
  \right]
\end{equation}
where $ M_\alpha $ is a $4 \times 4 $ matrix,
      $ M_\beta $ is  a $4 \times 2 $ matrix,
and   $ \tilde{M}_m $ is a $2 \times 2 $  diagonal matrix.
In particular, the matrix elements 
$ {M_\alpha}_{ij} $ 
for 
$ i,j =1,2,3 $
are zero. 
The values of the other matrix elements 
can be taken to be arbitrary.
The next step is to block-diagonalize this mass matrix as follows;
\begin{eqnarray}
&&  \left[
        \matrix{
        \epsilon M_\alpha  & \epsilon M_\beta \cr 
        \epsilon M_\beta^{\dag} & \tilde{M}_m
        }
  \right]
 \ \   =   \\
&&  \left[
        \matrix{
        1 & \epsilon P \cr 
        - \epsilon P^{T} & 1
        }
\right]
\left[
        \matrix{
        Q & 0 \cr 
        0 & \tilde{M}_m
        }
  \right]
\left[
        \matrix{ 1 & - \epsilon P \cr 
        \epsilon P^{T} & 1
        }
\right] \,.       \nonumber
\end{eqnarray}
Here the  $4\times 2$ matrix $P$ and the light neutrino mass
matrix $Q$ are given by
\begin{eqnarray}
 P &=& M_\beta \tilde{M}_m^{-1}   \\
 Q &=& \epsilon M_\alpha - \epsilon^2 P \tilde{M}_m P^{T}
    = \epsilon M_\alpha 
     - \epsilon^2 M_\beta \tilde{M}_m^{-1} M_\beta^T \nonumber \,.
\end{eqnarray}
Finally, 
the light neutrino mass matrix $Q$ is diagonalized
by a $4 \times 4 $ unitary rotation matrix $U$.
The mass eigenstates (physical states) are given by 
\begin{equation}
\Psi_P
=
\left[ \matrix{  U & 0 \cr   
                 0 & 1  }
\right]
\cdot
\left[  \matrix{  1    & - \epsilon P \cr  
        \epsilon P^{T} & 1}
\right]
\cdot
\left[  \matrix{
        1 & 0 \cr 
        0 & R }
\right]
\cdot
 \left[ \matrix{ \psi_{(e, \mu, \tau),l} \cr   
                  \psi_{(e, \mu, \tau),S}         }
 \right]
\end{equation}

Let us now determine the masses of six physical neutrinos.
The mass matrix relevant for heavy neutrinos is $\tilde{M}_m$. 
The nonzero values of the matrix elements are of order $M$.
And the physical neutrino fields are given by
\begin{eqnarray}
\nu_{5} & = & \sum_{\alpha = e , \mu , \tau}
    \epsilon \left( P_{\alpha , 1} \nu_{\alpha ,l}
                      +P_{4,1} R_{1, \alpha} \nu_{\alpha ,S} \right)
    + R_{2,\alpha} \nu_{\alpha ,S}  \nonumber \\
  &\simeq&   R_{2,\alpha} \nu_{\alpha ,S}  \nonumber \\
\nu_{6} & = & \sum_{\alpha = e , \mu , \tau}
    \epsilon \left( P_{\alpha , 2} \nu_{\alpha ,l}
                      +P_{4,2} R_{1, \alpha} \nu_{\alpha ,S} \right)
    + R_{3,\alpha} \nu_{\alpha ,S}  \\
 &\simeq&  R_{3,\alpha} \nu_{\alpha ,S}  \nonumber \,.
\end{eqnarray}
The masses of the light neutrinos come from diagonalizing mass matrix 
$Q$ given by
\begin{equation}
Q = \epsilon M_\alpha 
- \epsilon^2 M_\beta \tilde{M}_m^{-1} M_\beta^T \,.
\end{equation}
Neglecting $\epsilon^2$ term,  the most general matrix is
\begin{equation}
Q = \epsilon M_\alpha
   = \left[ \matrix{
            0 & 0 & 0 & a \cr
            0 & 0 & 0 & b \cr
            0 & 0 & 0 & c \cr
            a & b & c & 0 }
     \right] \,.
\end{equation}
The eigenvalues of $Q$ are two zeros
and $ \pm \sqrt{a^2 + b^2 + c^2 } $.
These two-fold degeneracies  are lifted by the 
$\epsilon^2$ term 
$- \epsilon^2 M_\beta \tilde{M}_m^{-1} M_\beta^T$.
Therefore, the light neutrinos are, in general,
two with mass $\epsilon^2 M$ and 
two with mass $\epsilon M$,
their mass difference being of order $\epsilon^2 M$.

When we neglect the $\epsilon^2$ term in the mass matrix,
the neutrino mass eigenstates are
\begin{equation}
\nu_{3,4} = \frac{1}{\sqrt{2}}
\left( \frac{ a \nu_e + b \nu_\mu + c \nu_\tau}
            {\sqrt{a^2 + b^2 + c^2}}
 \pm \nu_{S} \right)
\end{equation}
where $\nu_S = R_{1\alpha} \nu_{\alpha, S}$
is the massless component of the sterile 
neutrino mass matrix.  Other two neutrino states
(whose masses are of order $\epsilon^2 M$)
are orthogonal combinations of $\nu_{3,4}$.
Including the $\epsilon^2$ term in the light neutrino mass matrix $Q$,
it will give an order of $\epsilon$ to the neutrino mixing.
Therefore, we know that  the 
lightest pair of the neutrinos with mass $\epsilon^2 M$
are most likely active neutrinos.
However, 
the compositions of the $\nu_3$ and $\nu_4$ are 
dependent on the form of the mass matrix.
For example, 
let $ a=b=c$,
then the electron, muon and tau neutrino component 
in $\nu_{3,4}$ is $1/6$, respectively.
The other half of these neutrinos is the sterile neutrino.
As another example, consider $ c=0$ and $a=b$,
then the electron and muon neutrino component in 
$\nu_{3,4}$ is $1/4$, respectively, 
and the other half of $\nu_{3,4}$ is sterile.

We summarize the neutrino mass eigenvalues 
and their mass-squared differences which are relevant
to the known neutrino experiments for our discussions.
The neutrino masses are determined as follows:
\begin{eqnarray} \label{mns}
m_{\nu_1} &\simeq& m_{\nu_2} \simeq \epsilon^2 M  \nonumber \\
m_{\nu_3} &\simeq& m_{\nu_4} \simeq \epsilon M           \\
m_{\nu_5} &\simeq& m_{\nu_6} \simeq M \,.        \nonumber
\end{eqnarray}
Two lightest neutrinos, $\nu_1$ and $\nu_2$, 
are mostly  active neutrinos.
Two medium-weighted neutrinos, $\nu_3$ and $\nu_4$, are 
almost equal combinations of active and sterile neutrinos.
Two heaviest, $\nu_5$ and $\nu_6$, are 
almost sterile neutrinos.
Therefore, the mass-squared differences are given by
\begin{eqnarray}
\Delta m_{21}^2 
                &=& (m_{\nu_2} + m_{\nu_1})(m_{\nu_2} - m_{\nu_1}) 
                \simeq  \epsilon^4 M^2  \nonumber \\
\Delta m_{43}^2 
                &=& (m_{\nu_4} + m_{\nu_3})(m_{\nu_4} - m_{\nu_3}) 
                \simeq  \epsilon^3 M^2  \\
\Delta m_{42}^2 
                &=& (m_{\nu_4} + m_{\nu_2} )( m_{\nu_4} - m_{\nu_2} )
                \simeq  \epsilon^2 M^2  \nonumber\,,
\end{eqnarray}
and 
$\Delta m_{42}^2
\simeq \Delta m_{41}^2
\simeq \Delta m_{32}^2
\simeq \Delta m_{31}^2$.

\section{Determination of $\epsilon$, $M$  and neutrino masses}

As shown above, the singular seesaw mechanism has three $\Delta m^2$ scales 
which provide a possibility of explaining the three known experiments.
It is, however,  a priori uncertain  whether three experimental data 
(1), (2) and (3) can be accommodated simultaneously  since the three 
$\Delta m^2$ are parametrized by only two numbers $\epsilon$ and $M$.
We wish to see if the scheme (B) in Eq.~(\ref{AB}) can indeed be realized.  
Note first that the large mixing required for the atmospheric 
neutrino oscillation is built in the singular seesaw mechanism. 
That is, a combination of active neutrinos and a sterile neutrino have the 
maximal mixing to yield degenerate neutrinos $\nu_3, \nu_4$.
Therefore, the atmospheric neutrino problem can be explained by the large
mixing $\nu_\mu \leftrightarrow \nu_s$ oscillation with the 
mass-squared difference $\Delta m_{43}^2 \simeq \epsilon^3 M^2$.
The solar neutrino problem is then to be solved by the 
$\nu_e\leftrightarrow \nu_\tau$ oscillation with the smallest 
mass-squared difference $\Delta m_{21}^2 \simeq \epsilon^4 M^2$.  
As a consequence, the $\nu_\mu \leftrightarrow \nu_e$ oscillation occurs 
automatically with the largest mass-squared difference 
$\Delta m_{42}^2 \simeq \epsilon^2 M^2 $, which may yield observable 
signals in the $\nu_\mu \leftrightarrow \nu_e$ oscillation experiments.  

Let us now look for the parameter region of ($\epsilon$, $M$)
determined by the neutrino experiments, that is, 
$\epsilon^4 M^2=\Delta m^2_{\rm solar}$, 
$\epsilon^3 M^2=\Delta m^2_{\rm atm}$ and/or  
$\epsilon^2 M^2=\Delta m^2_{\rm LSND}$. 
From the first two identities based on the MSW solution to the solar 
neutrino problem and the atmospheric neutrino oscillation, one finds the
region of a crescent shape inside 
$ \epsilon = (0.43 \sim 9.6)\times 10^{-3}$ and 
$M=(0.02 \sim 9.3)$ keV.  
Remarkably, this region give rise to $\Delta m^2_{42}=\epsilon^2 M^2$ 
in the sensitivity range of the  LSND and KARMEN experiments.
Imposing the LSND positive result, the allowed region of ($\epsilon$, $M$)
is further reduced, and in fact determined solely
by the solar neutrino and LSND data. 
The common region of ($\epsilon$, $M$) 
which can explain all three known experiments lies inside
\begin{equation} \label{it}
\epsilon = (1.1 \sim 6.4) \times 10^{-3} \,,  \quad
M = (0.086 \sim 1.3) \, {\mathrm{keV}}  \,,
\end{equation}
which is shown in Fig.~1.
From Eqs.~(\ref{mns}) and (\ref{it}), we find the following typical 
values for the neutrino masses; 
\begin{eqnarray} \label{masses}
&& m_{\nu_{1,2}} \sim 10^{-3} \, {\mathrm{eV}}  \nonumber \\
&& m_{\nu_{3,4}} \sim 1  \, {\mathrm{eV}}       \\
&& m_{\nu_{5,6}} \sim 1\, {\mathrm{keV}}  \,.       \nonumber
\end{eqnarray}

If the solar neutrino deficit is due to the vacuum oscillation,
the solar neutrino data and the atmospheric neutrino data yields
$\epsilon = \Delta m_{21}^2/\Delta m_{43}^2 \lesssim
3\times 10^{-7}$.
Then the heavy right-handed neutrinos ($\nu_{5,6}$) get mass 
$ 100\ {\rm MeV} \lesssim M \lesssim 50\ {\rm GeV}$, and 
$ 30\ {\rm eV} \lesssim m_{\nu_{\mu,s}} \lesssim 700\ {\rm eV}$.  
The right-handed neutrinos with such masses may overclose the universe 
as they cannot decay into the standard model particles.
The muon and light sterile neutrinos can be candidates for hot dark matter
satisfying the overclosure bound 
$\sum_i m_{\nu_i} \lesssim 94 h^2 \Omega_\nu\ {\rm eV}$.
But they are too heavy to provide a good fit for the structure formation
as $\Omega_\nu \lesssim 0.3$ is required in mixed dark matter models 
\cite{MDM,MDMa}.
Furthermore, one finds no region of 
($\epsilon$,$M$) which accommodates all the three neutrino experiments.
Therefore, our model disfavors the possibility of solving the solar 
neutrino problem in terms of the vacuum oscillation.

It follows from the result (\ref{masses}) that 
almost degenerate muon and sterile neutrinos can
be good candidates for hot dark matter \cite{MDMa}, 
and the heavy sterile neutrino can be a candidate for 
the warm dark matter.
Thus, the existence of hot dark matter desirable 
for structure formation in the mixed dark matter scenario
is a natural consequence of the singular seesaw mechanism 
under consideration.

\section{Discussions on the mass scale of the sterile neutrinos}

It seems unnatural to have sterile (right-handed)
neutrinos with mass scale 
$M \approx {\mathrm keV}$.  
In the usual seesaw mechanism its scale is about
$10^{12} \, {\mathrm GeV}$ or GUT scale.
Such a low mass scale of lepton number violation is not acceptable.  
In order to raise the lepton number violation scale, we introduce
the ``double seesaw mechanism'' in which extra  sterile neutrinos 
are needed in addition to the ordinary right-handed neutrinos.
A simple realization of the double seesaw mechanism can be 
found in grand unified theory (GUT) with intermediate step breakings.
The minimal (non-supersymmetric) grand unification model is ruled out
from the study of the gauge coupling evolution\cite{GKLGUT},
but non-minimal GUT models and supersymmeric GUT models 
can successfully accommodate the gauge coupling unification
and the absence of proton decay.  
As an example, let us consider a GUT model with
$E_6$ unification group.  It has five neutral particles
in each generations of the fermion {\bf 27} representation:
Two ({\bf 16} under SO(10)) of them are the neutrinos in the discussion,
other three neutrinos ({\bf 1} + {\bf 10} under SO(10))
are heavy neutrinos with GUT scale masses.
Suppose that the mass matrix of the active neutrinos, the right-handed 
neutrinos and the extra sterile neutrinos takes the form;
\begin{equation} \label{basicM}
{\cal M} =
\left[
\matrix{
0   & 0   & M_L \cr
0   & 0   & M_R \cr
M_L^T & M_R^T & M_S }
\right] \,.
\end{equation}
Here $M_L$ is the Dirac mass matrix originating from the electroweak symmetry
breaking, and $M_R, M_S$ are generated from a higher symmetry breaking.
Note that $M_{L}, M_R$ are $3 \times 9$ matrices, and $M_S$ is $9\times9$
matrix in the three generation model.
It should be mentioned that the mass matrix (\ref{basicM}) requires
fine-tunings which may not be a serious problem in supersymmetric theories.
Given the hierarchy $M_L << M_R << M_S$, the seesaw mechanism with the
ultra heavy neutrino masses $M_S$ gives rise to the $6\times6$ matrix, 
\begin{eqnarray} \label{Msub}
{\cal M}_{sub}
&=& -
\left[ \matrix{  M_L \cr \cr   M_R  }
\right]
\cdot
M_S^{-1}
\cdot
\left[ \matrix{  M_L^T & M_R^T  } \right]
\nonumber \\[2mm]
&=& -
\left[
\matrix{ M_L M_S^{-1} M_L^T   &  M_L M_S^{-1} M_R^T    \cr \cr
M_R M_S^{-1} M_L^T &  M_R M_S^{-1}  M_R^T }
\right]
\label{matrixLRS}
\end{eqnarray}
which can be identified with the matrix (\ref{firstM}) apart from the
upper-left corner.
Note that the singular seesaw mechanism requires the lower-right submatrix 
of Eq.~(\ref{Msub}) to be singular.  The  nonzero contribution in 
the upper-left part of the matrix (\ref{Msub}) 
is order of the solar neutrino mass scale $\epsilon^2 M \sim 10^{-3}$ eV which
does not alter our conclusion.

Following the discussion in the previous section,
we can find the typical scales of $M_R$ and $M_S$.
Namely, requiring $ \epsilon =M_L/M_R$, $M=M_R^2/M_S$, and 
$M_L \simeq 100$ GeV, one obtains
\begin{equation}
 M_R \approx 10^5 {\mathrm GeV} \,, \quad
 M_S \approx 10^{16} {\mathrm GeV} \,. 
\end{equation}
It is worth emphasizing that the heaviest mass scale $M_S$ coincides with
the conventional GUT scale.  On the other hand, the intermediate scale $M_R$ 
turns out to be considerably lower than the conventional scale $10^{12}$ GeV
desirable for the usual seesaw mechanism.
In GUT models, various intermediate scales of gauge symmetry breaking 
can be made consistent with the unification of gauge coupling constants.
In particular, such a low scale $M_R$ may be obtained by introducing
certain exotic particles to the GUT model \cite{LWFOOT}.

The scale $M_R$ may be related to physics of supersymmetry breaking.
In the supersymmetric standard model, supersymmetry breaking can be 
mediated either by gravitation or by gauge interactions \cite{GMSB}.  
The latter scheme provides a natural suppression
of flavor violation in the supersymmetric sector and yields distinctive
phenomenological and cosmological consequences \cite{zzz,xxx}.
The minimal type of such theories requires the existence of 
vector-like quarks and leptons (messengers) 
at the mass scale $(10^4 \sim 10^5)$ GeV.
This is a just right scale for $M_R$ under discussion.
Therefore, in the supersymmetric standard model with gauge-mediated 
supersymmetry breaking, it is conceivable that some messengers are
vector-like sterile neutrinos with the mass matrix 
given in Eq.~(\ref{basicM}).

\section{Conclusions}

Motivated by recent experimental evidences for nonzero neutrino masses and 
mixing, we have examined the consequences of the singular seesaw mechanism.
The three mass-squared scales required for simultaneously
explaining the solar and 
atmospheric neutrino anomalies, and the LSND data can be realized
if the Majorana mass matrix of right-handed neutrinos is rank-two.  
Without assuming any hierarchies in the Dirac and Majorana mass matrices,
three mass-squared values are found to be determined by
two mass parameters: the Dirac mass for the active neutrinos 
and the Majorana mass for heavy right-handed neutrinos.
The singular seesaw mechanism cannot accommodate simultaneously
the vacuum oscillation explanation of the solar neutrino deficit, and 
the atmospheric neutrino oscillation.
However, the MSW solution to the solar neutrino problem is consistent with
the model and the existence of the LSND mass scale is also explained.
The almost maximal mixing of a sterile neutrino with 
the muon neutrino (having the Dirac mass $m_{\nu_{3,4}} \sim 1$ eV)
explains the atmospheric neutrino anomaly, and  the mixing of the 
electron and tau neutrino explains the solar neutrino anomaly (having
the lightest mass $m_{\nu_{1,2}} \sim 10^{-3}$ eV).  
Two massive right-handed neutrinos turn out to be rather light 
(having the Majorana mass $m_{\nu_{5,6}} \sim 1$ keV). 
We stress that the existence of hot dark matter (consists of $\nu_{\mu,s}$)
desirable for the structure formation of the universe is a natural
consequence of our scheme.  In addition, warm dark matter can be provided by
the heavy right-handed neutrinos.
We have introduced the double seesaw mechanism in which the two low mass
scales, $m_{\nu_{3,4}}$ and $m_{\nu_{5,6}}$ are generated by the weak scale
$M_L \sim 100$ GeV  and an intermediate scale $M_R \sim 10^5$ GeV 
together with the usual grand unification scale $M_S \sim 10^{16}$ GeV.
A candidate for the intermediate scale $M_R$ can be found in the GUT models 
with an intermediate step breaking, or in gauge-mediated  
supersymmetry breaking models.

\acknowledgments  
UWL thanks KIAS for the kind hospitality during his visit.
EJC would like to thank Alexei Smirnov
for various comments on this work, and ICTP for its 
hospitality during the Extended workshop on Highlights in Astroparticle
Physics.  EJC  is supported by Non-Directed Research Fund of 
Korea Research Foundation, 1996.

\end{multicols}
\vspace*{-2.0in}
\centerline{\epsfig{file=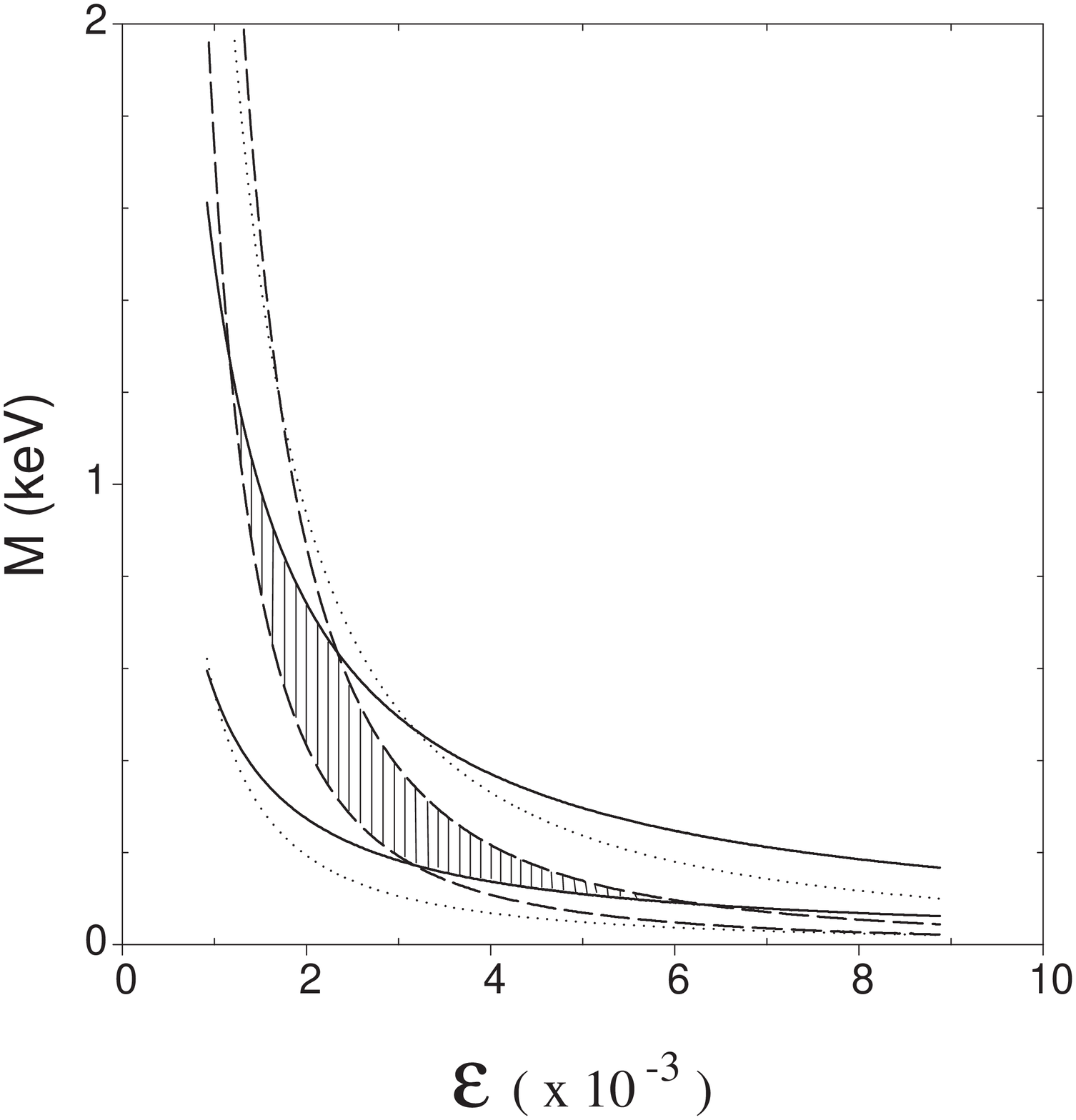,width=6.5in}}
 Fig.~1. 
The regions between the dashed lines, the dotted lines, and the solid lines
are allowed by the solar neutrino data, the atmospheric neutrino data, and 
the LSND data, resepectively.  The shaded region accomodates all the 
three neutrino experiments.
\end{document}